\documentclass[12pt]{article}
\usepackage{epsfig,amsmath,amsfonts,amssymb,amstext,afterpage,psfrag}
\usepackage{slashed}
\usepackage[square,comma,sort&compress,numbers]{natbib}
\usepackage{xcolor}
\allowdisplaybreaks
\long\def\symbolfootnote[#1]#2{\begingroup%
\def\thefootnote{\fnsymbol{footnote}}\footnote[#1]{#2}\endgroup}

\def\spose#1{\hbox to 0pt{#1\hss}}
\def\lsim{\mathrel{\spose{\lower 3pt\hbox{$\mathchar"218$}}
 \raise 2.0pt\hbox{$\mathchar"13C$}}}
\def\gsim{\mathrel{\spose{\lower 3pt\hbox{$\mathchar"218$}}
 \raise 2.0pt\hbox{$\mathchar"13E$}}}

\catcode`@=11
\def\@citex[#1]#2{%
  \if@filesw\immediate\write\@auxout{\string\citation{#2}}\fi
  \def\@citea{}\@cite{\@for\@citeb:=#2\do
    {\@citea\def\@citea{,\penalty\@m}\@ifundefined
      {b@\@citeb}{{\bf ?}\@warning
{Citation `\@citeb' on page \thepage \space undefined}}%
      \hbox{\csname b@\@citeb\endcsname}}}{#1}}
\def\citer{\@ifnextchar [{\@tempswatrue\@citexr}{\@tempswafalse\@citexr[]}}
  \def\@citexr[#1]#2{%
    \if@filesw\immediate\write\@auxout{\string\citation{#2}}\fi
    \def\@citea{}\@cite{\@for\@citeb:=#2\do
      {\@citea\def\@citea{--\penalty\@m}\@ifundefined
{b@\@citeb}{{\bf ?}\@warning
{Citation `\@citeb' on page \thepage \space undefined}}%
\hbox{\csname b@\@citeb\endcsname}}}{#1}}

\begin{document}

\begin{titlepage}

\begin{flushright}
{\small
LMU-ASC~02/26\\
MBI-ML-26-02\\
April 2026
}
\end{flushright}

\vspace{0.5cm}
\begin{center}
{\Large\bf \boldmath                                               
Comment on ``Primary Dimensions''
\unboldmath}
\end{center}

\vspace{0.5cm}
\begin{center}
{\sc G. Buchalla$^a$, O. Cat\`a$^a$, A. Celis$^a$ and C. Krause$^b$} 
\end{center}

\vspace*{0.4cm}

\begin{center}
$^a$Ludwig-Maximilians-Universit\"at M\"unchen, Fakult\"at f\"ur Physik,\\
Arnold Sommerfeld Center for Theoretical Physics, 
D--80333 M\"unchen, Germany\\
\vspace*{0.2cm}
$^b$Marietta Blau Institute for Particle Physics (MBI Vienna),\\
Austrian Academy of Sciences (\"OAW), Austria
\end{center}

\vspace{1.5cm}
\begin{abstract}
\vspace{0.2cm}\noindent
We show that the concept of {\it primary dimensions\/},
first introduced in \cite{Gavela:2016bzc} as an organizing
principle for chiral Lagrangians, is inconsistent.
Although this had been pointed out already
in \cite{Buchalla:2016sop}, the notion of primary dimensions
has re-appeared in recent literature.
We briefly comment on the proper power counting for such effective
field theories, including the electroweak chiral Lagrangian
with a light Higgs, which is based on chiral dimensions, equivalent
to the counting of loop orders.
\end{abstract}

\vfill

\end{titlepage}

The article \cite{Gavela:2016bzc} presents a general
discussion of power counting in effective field theories (EFTs),
in particular for the electroweak chiral Lagrangian including
a light Higgs, termed HEFT in \cite{Gavela:2016bzc}, and casts doubt
on the concept of chiral dimensions.
Except for the discussion of HEFT in sec. 5, ref.~\cite{Gavela:2016bzc}
essentially gives a review of known material. 
For instance, a master formula can be written that determines the
(parametric) size of the coefficient of a term in a low-energy EFT
with cutoff $\Lambda$, which contains scalar fields $\phi$,
gauge fields $A$, fermions $\psi$, gauge couplings $g$, Yukawa couplings $y$,
scalar self couplings $\lambda$ and derivatives~$\partial$.  
This master formula, quoted in eq. (23) of ref.~\cite{Gavela:2016bzc},
\begin{equation}\label{eftmaster2}
\frac{\Lambda^4}{16\pi^2} \left[\frac{\partial}{\Lambda} \right]^{N_p}
\left[\frac{4\pi\phi}{\Lambda} \right]^{N_\phi}
\left[\frac{4\pi A}{\Lambda} \right]^{N_A}
\left[\frac{4\pi\psi}{\Lambda^{3/2}} \right]^{N_\psi}
\left[\frac{g}{4\pi} \right]^{N_g}
\left[\frac{y}{4\pi} \right]^{N_y}
\left[\frac{\lambda}{16\pi^2} \right]^{N_\lambda}
\end{equation}
has been previously known in the
literature (see e.g. \cite{Buchalla:2013eza} and references therein).
The master formula in (\ref{eftmaster2}) can be shown to follow just
from the counting of canonical ($d_c$) and chiral dimensions ($d_\chi$),
as we argued in \cite{Buchalla:2016sop}.
In particular, in \cite{Buchalla:2016sop} it was shown that the powers
of $\Lambda$ and $4\pi$ in (\ref{eftmaster2}) are given by
$\Lambda^{4-d_c} (4\pi)^{d_c-d_\chi -2}$.
The assignment of chiral dimension $d_\chi$ is 1 for each derivative,
weak coupling and fermion bilinear, and 0 for bosons. 
The total chiral dimension $d_\chi$ of a term in the Lagrangian determines
its loop order $L$ through $d_\chi = 2L + 2$ \cite{Buchalla:2013eza}.
The master formula and chiral counting are thus perfectly consistent.

The genuinely new material of \cite{Gavela:2016bzc} is contained
in its sec. 5. Sec. 5 introduces
the concept of {\it primary dimensions\/} as a new organizing principle
for chiral Lagrangians. As we will elaborate here, 
the claims presented in sec. 5 are incorrect. 
As indicated in the Introduction of ~\cite{Gavela:2016bzc},
the discussion in sec. 5 was motivated by several of our previous 
papers \cite{Buchalla:2013rka,Buchalla:2013eza,Buchalla:2014eca,Buchalla:2015wfa,Buchalla:2015qju},
in which the systematics of power counting for the Higgs-electroweak chiral 
Lagrangian had already been addressed in great detail, and had been shown
to conform with the usual rules of chiral counting \cite{Buchalla:2013eza}.

In sec. 5 of \cite{Gavela:2016bzc}, 
$v=246\,{\rm GeV}$ denotes the electroweak
vacuum expectation value, and $f$ the scale of some new Higgs dynamics,
with $v < f$. This notation is widely used and uncontroversial here.
The three electroweak Goldstone bosons $\Pi^a$ can be parametrized
by the field $U=\exp(2 i\Pi/v)$, $\Pi=\Pi^a T^a$, with $T^a$ the generators 
of $SU(2)$. The field $U$ transforms linearly under $g_L\in SU(2)_L$,
$g_R\in SU(2)_R$ as $U\to g_L U g^\dagger_R$.

When gauged under the electroweak symmetry $SU(2)_L\otimes U(1)_Y$,
the $U$-field kinetic term
\begin{equation}\label{lukin}
{\cal L}_U=\frac{v^2}{4}\, {\rm tr} D_\mu U^\dagger D^\mu U\, ,\qquad
D_\mu U =\partial_\mu U + i g W_\mu U -i g' B_\mu U T_3
\end{equation}
yields in particular the mass term of the electroweak gauge bosons.
In the renormalizable Standard Model with the usual Higgs scalar $h$,
the term ${\cal L}_U$ comes with a factor $(1+h/v)^2$. In the electroweak
chiral Lagrangian with a light Higgs, this factor is generalized to a
function $(1+F(h/v))$. All of this is well known in the literature,
see e.g. \cite{Feruglio:1992wf} and \cite{Contino:2010rs}.

In sec. 5 of \cite{Gavela:2016bzc} the Goldstone field $U$ 
and the kinetic term are replaced by
\begin{equation}\label{lufkin}
{\cal L}_{U,f}=\frac{f^2}{4}\, {\rm tr} D_\mu U^\dagger D^\mu U\, ,\qquad
U=\exp(2 i\Pi/f)
\end{equation}
substituting the new-physics scale $f$ for the electroweak scale $v$.
As a consequence, the $W$-boson mass would then turn out to be
$M_W=gf/2$, rather than the correct value $M_W=g v/2$.

In order to avoid this problem, the gauge couplings in the covariant
derivative for $U$ are multiplied by a factor of $v/f$ 
(eq. (80) of \cite{Gavela:2016bzc}). That is, focusing on the  
$SU(2)_L$-part, \cite{Gavela:2016bzc} imposes
\begin{equation}\label{dmugvf}
D_\mu U =\partial_\mu U + i g \frac{v}{f} W_\mu U
\end{equation} 
However, this is inconsistent with gauge invariance:
The $SU(2)_L$-doublet fermions have a covariant derivative
$D_\mu \psi_L=(\partial_\mu + i g W_\mu)\psi_L$,
defining the weak coupling $g$ of fermions to $W$ in the usual way.
For the Yukawa term $\bar\psi_L U\psi_R$, eq. (81) of \cite{Gavela:2016bzc},
to be gauge invariant under $SU(2)_L$, the gauge transformation of $U$
has to match the one of $\psi_L$. Consequently, one must have
$D_\mu U =\partial_\mu U + i g W_\mu U$ under $SU(2)_L$, in contradiction to 
the modification $g\to g v/f$ in eq. (80) of \cite{Gavela:2016bzc}.

Apparently, the motivation for replacing $v\to f$ in the Goldstone field
$U$ in \cite{Gavela:2016bzc} has been to enable a series expansion
of the effective Lagrangian in inverse powers of the new-physics scale $f$.
The remainder of sec. 5 of \cite{Gavela:2016bzc} is devoted to describing
such an expansion, resulting in the definition of primary dimensions
(defined for a given operator as the smallest operator mass dimension
of the terms resulting from its series expansion in $1/f$). 
As we have shown here, already the starting point 
for this procedure, the substitution $v\to f$ in (\ref{lufkin}), is invalid. 
For this reason, sec. 5 of \cite{Gavela:2016bzc} has no basis.

Recently, the notion of primary dimensions, along the lines
of~\cite{Gavela:2016bzc}, has been employed in~\cite{Brivio:2025yrr}
as a possible option for the operator expansion in HEFT,
in particular in section 5.2 and eq. (5.68).
The above conclusions also apply to this case.

A more detailed discussion and further references
can be found in \cite{Buchalla:2016sop}.
It is reviewed there how chiral dimensions 
define an unambiguous power counting to systematically build an 
electroweak chiral EFT with a light Higgs, rendering the EFT renormalizable
order by order in the operator expansion.
The application of chiral dimensions is illustrated with various
examples, including chiral perturbation theory with pions and photons
as an instructive analogy to HEFT.



\end{document}